\documentclass[10pt,doublecolumn,twoside,journal]{IEEEtran}
\IEEEoverridecommandlockouts
\usepackage{CJK}
\usepackage{subfigure}
\usepackage{colortbl}
\usepackage{bm}
\usepackage{graphicx}  
\usepackage{url}       

\usepackage{amsmath}
\usepackage{cite}

\usepackage{amsfonts,amssymb}

\usepackage{stfloats}

\usepackage{cases}
\usepackage{algorithm}
\usepackage{multirow}
\usepackage{algorithmic}
\usepackage{float}

\newcommand{\ls}[1]
    {\dimen0=\fontdimen6\the\font
     \lineskip=#1\dimen0
     \advance\lineskip.5\fontdimen5\the\font
     \advance\lineskip-\dimen0
     \lineskiplimit=.9\lineskip
     \baselineskip=\lineskip
     \advance\baselineskip\dimen0
     \normallineskip\lineskip
     \normallineskiplimit\lineskiplimit
     \normalbaselineskip\baselineskip
     \ignorespaces
    }

\begin{document}
%
\title{Robust Beamforming for Magnetic Induction Based Underground Emergency Communications}
\author{Jianyu~Wang,~\IEEEmembership{Member,~IEEE,}
        Tianrui~Hou,
        Wenchi~Cheng,~\IEEEmembership{Senior~Member,~IEEE,}

        and~Hailin~Zhang,~\IEEEmembership{Member,~IEEE}

\thanks{

Jianyu Wang, Tianrui Hou, Wenchi Cheng, and Hailin Zhang are with State Key Laboratory of Integrated Services Networks, Xidian University, Xi'an, 710071, China (e-mails: wangjianyu@xidian.edu.cn; houtianrui@stu.xidian.edu.cn; wccheng@xidian.edu.cn; hlzhang@xidian.edu.cn)

}
\vspace{-20pt}
}

\maketitle

\thispagestyle{empty}

\begin{abstract}
Magnetic induction (MI) communication is an effective underground emergency communication technique after disasters such as landslides, mine collapses, and earthquakes, due to its advantages in mediums such as soil, concrete, and metals.
Based on channel state information (CSI), magnetic beamforming can significantly improve the performance of MI communication.
However, in post-disaster underground communication, channel estimation may suffer from errors due to factors such as complex environmental interferences.
Taking channel estimation error into account, we formulate a beamforming optimization problem for multi-user MI underground emergency communications, which aims to minimize the power consumption under the constraints of sum rate and signal to interference plus noise ratio (SINR) of each user.
Based on the worst-case optimization criterion and the S-procedure, the non-convex optimization problem is transformed into convex and solved. Numerical results show that the proposed robust beamforming scheme can effectively  enhance communication reliability and effective throughput in the presence of channel estimation errors.
\end{abstract}

\begin{IEEEkeywords}
Magnetic induction (MI) communication, robust beamforming, underground emergency communications.
\end{IEEEkeywords}

\section{introduction}
\IEEEPARstart{D}{ue} to the advantages in special mediums such as soil, concrete, and metals, magnetic induction (MI) communications have gained substantial attention in recent years and are considered to support novel and important applications in future underground emergency rescues.
Magnetic beamforming, which utilizes multiple coils to constructively combine the induced magnetic fields, has emerged as an effective approach to enhance the performance of MI communications.

Recent studies have explored various aspects of this technique in different system configurations.
For multiple input multiple output (MIMO) wireless power transfer (WPT), the magnetic beamforming and the channel estimation method were investigated in~\cite{Gang_Yang_TSP}.
Also, our recent works demonstrated that integrating magnetic beamforming with other techniques, such as Multi-frequency Resonating Compensation (MuReC) coil arrays~\cite{JSAC_2022} and orthogonal frequency division multiplexing (OFDM)~\cite{ICCC_2021} can significantly enhance achievable data rates for MI communications. Nevertheless, above works assumes perfect channel state information (CSI) at the transmitter, which is usually difficult to realize in emergency rescue scenarios due to factors such as complex environmental interferences. Considering channel estimation errors, random beamforming~\cite{Random_Beamforming} and robust beamforming~\cite{Yixin_Zhang_TSP} methods were proposed for WPT, which focus on enhancing the power transfer reliability. Different from WPT, in MI based underground emergency communication, it is essential to guarantee the signal to interference plus noise ratio (SINR) of each user and network throughput. To the best of our knowledge, magnetic beamforming for MI based underground emergency communication with imperfect CSI has not been discussed.

In this paper we propose a robust magnetic beamforming scheme to enhance reliability and effective throughput in MI based multi-user underground emergency communications. Considering imperfect CSI, we formulate a beamforming optimization problem to minimize power consumption while ensuring both the total sum rate and per-user SINR. Using worst-case optimization and the S-procedure, the original non-convex problem is transformed into a convex form and solved. Numerical results verify our analysis and show the superiority of the proposed scheme in the case of imperfect CSI.

The remainder of this paper is structured as follows. In Section~II, we formulate the system model for MI based multi-user MIMO underground emergency communication network. Also, considering the imperfect CSI, a magnetic beamforming  problem is formulated. In Section~III, the original problem is converted into convex and a robust magnetic beamforming scheme is proposed. We conduct numerical simulations in Section~IV and conclude in Section~V.
\section{System Model and Problem Formulation}
We consider an MI based multi-user underground emergency communication network, where the above-ground emergency access point (EAP) transmits information to $N$ underground trapped users (TUs) simultaneously $(N\leq K)$. The EAP is equipped with $K$ coils and each TU is equipped with single coil. The operating frequency-band is denoted by $\mathcal B=\{f|f_c-B/2\leq f\leq f_c+B/2\}$, where $f$ denotes frequency, $B$ denotes bandwidth, and $f_c$ denotes center frequency.
We denote by $Z_{a,k}(f)$ the impendence of the $k$-th coil of the EAP and $Z_{u,n}(f)$ the impendence of the $n$-th TU.
The mutual inductance between the $p$-th coil and the $q$-th coil of EAP is denoted by $\tilde{M}_{pq}(f)$. Then, we can write the impendence matrix of EAP, denoted by $\mathbf{Z}_{a}(f)$, as follows~\cite{Random_Beamforming}:
\begin{equation}\label{eq:Z_t_n}
\mathbf{Z}_{a}(f)\!\!=\!\!\!\left[\!\begin{array}{cccc}
\!Z_{a,1}(f)\!\! & \!\!\!j 2\pi f \tilde M_{{1,2}}(f)\!\! & \!\!\!\!\!\!\cdots\!\!\!\! & \!\!\!\!j 2\pi f \tilde M_{1, K}(f)\!\! \\
\!j 2\pi f \tilde M_{{2,1}}(f)\!\! & \!\!\!Z_{a,2}(f)\!\! & \!\!\!\!\!\!\cdots\!\!\!\! & \!\!\!\!j 2\pi f \tilde M_{2, K}(f)\!\! \\
\!\vdots\!\! & \!\!\vdots\!\! & \!\!\!\ddots\!\! & \!\!\!\!\!\!\vdots\!\!\!\! \\
\!j 2\pi f \tilde M_{K, 1}(f)\!\! & \!\!j 2\pi f \tilde M_{K, 2}(f)\!\! & \!\!\!\!\!\!\cdots\!\!\!\! & \!\!\!\!Z_{a,K}(f)\!\!
\end{array}\!\!\!\right]\!\!,
\end{equation}
where diagonal and off-diagonal entries are coil impendence of EAP and mutual inductances among coils of EAP, respectively.
We denote by $M_{kn}(f)$ the mutual inductance between the $k$-th coil of EAP and the $n$-th TU. Then, the mutual inductance matrix between EAP and all TUs, denoted by $\mathbf{M}(f)$, can expressed based on $\mathbf{M}(f)$ as follows~\cite{Random_Beamforming}:
\begin{equation}\label{eq:M}
\mathbf{M}(f)=\!\!\!\left[\!\begin{array}{cccc}
\!M_{1,1}(f)\!\! & \!\!\!M_{2,1}(f)\!\! & \!\!\!\!\!\!\cdots\!\!\!\! & \!\!\!\!M_{K,1}(f)\!\! \\
\!M_{1,2}(f)\!\! & \!\!\!M_{2,2}(f)\!\! & \!\!\!\!\!\!\cdots\!\!\!\! & \!\!\!\!M_{K,2}(f)\!\! \\
\!\vdots\!\! & \!\!\vdots\!\! & \!\!\!\ddots\!\! & \!\!\!\!\!\!\vdots\!\!\!\! \\
\!M_{1,N}(f)\!\! & \!\!\!M_{2,N}(f)\!\! & \!\!\!\!\!\!\cdots\!\!\!\! & \!\!\!\!M_{K,N}(f)\!\!
\end{array}\!\!\!\right]\!\!.
\end{equation}
The mutual inductance between the $m$-th TU and the $v$-th TU is denoted by $\hat{M}_{mv}(f)$. Then, we can write the impendence matrix of TUs, denoted by $\mathbf{Z}_{u}(f)$, as follows~\cite{Random_Beamforming}:
\begin{equation}\label{eq:Z_t_n}
\mathbf{Z}_{u}(f)\!\!=\!\!\!\left[\!\begin{array}{cccc}
\!Z_{u,1}(f)\!\! & \!\!\!j 2\pi f \hat M_{{1,2}}(f)\!\! & \!\!\!\!\!\!\cdots\!\!\!\! & \!\!\!\!j 2\pi f \hat M_{1, N}(f)\!\! \\
\!j 2\pi f \hat M_{{2,1}}(f)\!\! & \!\!\!Z_{u,2}(f)\!\! & \!\!\!\!\!\!\cdots\!\!\!\! & \!\!\!\!j 2\pi f \hat M_{2, N}(f)\!\! \\
\!\vdots\!\! & \!\!\vdots\!\! & \!\!\!\ddots\!\! & \!\!\!\!\!\!\vdots\!\!\!\! \\
\!j 2\pi f \hat M_{N, 1}(f)\!\! & \!\!j 2\pi f \hat M_{N, 2}(f)\!\! & \!\!\!\!\!\!\cdots\!\!\!\! & \!\!\!\!Z_{u,N}(f)\!\!
\end{array}\!\!\!\right]\!\!,
\end{equation}
where diagonal and off-diagonal entries are coil impendence of TUs and mutual inductances among coils of TUs, respectively.

Let $x_n(t)=\sum_{w=0}^{W-1}S_{n,w}g(t-wT_S)e^{j2\pi f_c(t-wT_S)}$ denote the transmit signal from the EAP to the $n$-th TU, where $W$ is the frame length, $S_{n,w}\sim\mathcal C \mathcal N(0,1)$ is the $w$-th symbol corresponding to the $n$-th TU, $T_S$ is the symbol period, and $g(t)$ is the base-band pulse with the bandwidth of $B=1/T_S$. It is assumed that for $\forall w_1,w_2\in\{0,....,W-1\}$ and $\forall n_1,n_2\in\{1,...,N\}$, where $w_1\ne w_2$ and $n_1\ne n_2$, $S_{n_1,w_1}$ and $S_{n_2,w_2}$ are independent. The energy consumption of $g(t)$ on a unit resistance is normalized to 1, that is, $\int|g(t)|^2dt\!=\!1$. Then, the average energy consumption of $x_n(t)$ on a unit resistance can be given by $\mathbb{E}\left\{\int|x_n(t)|^2dt\right\}=W$, where $\mathbb E\{\cdot\}$ denotes the operation of taking the expectation.
The transmit signals are mapped to coil currents of EAP. The current corresponding to the $k$-th coil of EAP, denoted by $i_{a,k}(t)$, is $i_{a,k}(t)=\sum_{n=1}^{N}\tilde I_{a,nk}x_{n}(t)$, where $\tilde I_{a,nk}$ is the magnetic beamforming element corresponding to the $n$-th TU and the $k$-th coil of EAP. Then, we can obtain the Fourier transform of $i_{a,k}(t)$, denoted by $I_{a,k}(f)$, as $I_{a,k}(f)=\sum_{n=1}^{N}\tilde I_{a,nk}X_{n}(f)$.
Based on this expression, the current vector of EAP in the frequency-domain, denoted by $\mathbf{I}_{a}(f)$, can be given by $\mathbf{I}_{a}(f)=\sum_{n=1}^{N}\mathbf{\tilde I}_{a,n}X_n(f)$,
where $\mathbf{\tilde I}_{a,n}=[\tilde I_{a,n1},\tilde I_{a,n2},...,\tilde I_{a,nK}]^T$ is the beamforming vector of EAP. Let $\mathbf{I}_{u}(f)=[I_{u,1}(f),I_{u,2}(f),...,I_{u,N}(f)]^T$ and $\mathbf{V}_{a}(f)=[V_{a,1}(f),V_{a,2}(f),...,V_{a,K}(f)]^T$ denote the current vector of TUs and the voltage vector of EAP in the frequency-domain, respectively, where $I_{u,n}(f)$ represents the frequency-domain current of the $n$-th TU and $V_{a,k}(f)$ represents the frequency-domain voltage of the $k$-th coil of EAP. According to Faraday's electromagnetic induction law, we have
\begin{equation}\label{eq:I_uf}
\mathbf{Z}_{u}(f)\mathbf{I}_{u}(f)=j2\pi f\mathbf{M}(f)\sum_{n=1}^{N}\mathbf{\tilde I}_{a,n}X_n(f)
\end{equation}
and
\begin{equation}\label{eq:V_af}
\mathbf{V}_{a}(f)=\mathbf{Q}(f)\sum_{n=1}^{N}\mathbf{\tilde I}_{a,n}X_n(f),
\end{equation}
where $\mathbf{Q}(f)=\mathbf{Z}_{a}(f)+4\pi^2f^2\mathbf{M}(f)^H\mathbf{Z}_{u}(f)^{-1}\mathbf{M}(f)$.
Based on~\eqref{eq:I_uf}, we can obtain the frequency-domain receive signal of the $n$-th TU, which is the voltage of the resistor and denoted by $Y_n(f)$, as follows:
\begin{equation}\label{eq:Y_uf}
Y_n(f)\!\!=\!\mathbf{H}_n(f)\mathbf{\tilde I}_{a,n}X_n(f)\!+\!\mathbf{H}_n(f)\!\!\sum_{u\ne n}^{N}\!\mathbf{\tilde I}_{a,u}X_u(f)\!+\!Z_n(f),
\end{equation}
where $Z_n(f)$ is noise, $\mathbf{H}_n(f)=j2\pi fR_{L,n}\mathbf{o}_n\mathbf{Z}_{u}(f)^{-1}\mathbf{M}(f)$, $\mathbf{o}_n$ is a $1\times N$ row selection vector to extract the $n$-th row of $\mathbf{Z}_{u}(f)^{-1}$, with the $n$-th element of $\mathbf{o}_n$ being 1 and the remaining elements being 0.
In this paper we focus on narrow-band transmission where $\mathbf{H}_n(f)\approx\mathbf{H}_n(f_c)$ and $\mathbf{Q}(f)\approx\mathbf{Q}(f_c)$ for $f\in \mathcal B$~\cite{JSAC_2022,Joint_Channel_Antenna_Modeling,Active_Relaying}. While our analysis considers flat fading channels, the method can be readily extended to frequency-selective channels using OFDM~\cite{ICCC_2021}. In the following we write $\mathbf{H}_n(f)$ and $\mathbf{Q}(f)$ as $\mathbf{H}_n$ and $\mathbf{Q}$, respectively.

Both $\mathbf{Q}$ and $\mathbf{H}_n$ are unknown for EAP at the beginning. To implement magnetic beamforming at the EAP side, both $\mathbf{H}_n$ and $\mathbf{Q}$ are desired to be estimated.
However, in practical post-disaster emergency rescue scenarios, due to factors such as limited power and complex environmental interferences, the estimation for $\mathbf{H}_n$ and $\mathbf{Q}$ may suffer from errors. The actual $\mathbf{H}_n$ and $\mathbf{Q}$ can be expressed as $\mathbf{H}_n=\mathbf{\bar H}_n+\mathbf{\Delta H}_n$ and $\mathbf{Q}=\mathbf{\bar Q}+\mathbf{\Delta Q}$, respectively,
where $\mathbf{\bar H}_n$ and $\mathbf{\bar Q}$ are estimation components, $\mathbf{\Delta H}_n$ and $\mathbf{\Delta Q}$ are error components. We use the norm-bounded model to characterize errors components and assume that they are bounded by possible values, that is, $\|\mathbf{\Delta H}_n\|_F\leq\xi_n$ and $\|\mathbf{\Delta Q}\|_F\leq\xi_q$~\cite{Yixin_Zhang_TSP}, where $\|\cdot\|_F$ denotes the operation of taking the Frobenius norm. Based on \eqref{eq:Y_uf}, the SINR of the $n$-th TU, denoted by $\gamma_n$, can be given by
\begin{equation}\label{eq:gamma_n}
\begin{split}
\gamma_{n}=\frac{ \mathbf{\tilde I}_{a,n}^H(\mathbf{\bar H}_n\!+\!\mathbf{\Delta H}_n)^H(\mathbf{\bar H}_n\!+\!\mathbf{\Delta H}_n)\mathbf{\tilde I}_{a,n}}{\sum_{u\ne n}^{N}\mathbf{\tilde I}_{a,u}^H(\mathbf{\bar H}_n\!+\!\mathbf{\Delta H}_n)^H(\mathbf{\bar H}_n\!+\!\mathbf{\Delta H}_n)\mathbf{\tilde I}_{u,n}\!+\!N_0}.
\end{split}
\end{equation}
Then, the sum rate of the network, denoted by $C_S$, can be expressed based on~\eqref{eq:gamma_n} as follows:
\begin{equation}\label{eq:C_s}
\begin{split}
C_S=\sum_{n=1}^{N}B\log_2(1+\gamma_n).
\end{split}
\end{equation}

Based on above analysis, we can obtain the time-domain current and voltage vectors as $\mathbf{i}_a(t)\!=\!\sum_{n=1}^{N}\mathbf{\tilde I}_{a,n}x_n(t)$ and $\mathbf{v}_a(t)\!=\!\mathbf{Q}\sum_{n=1}^{N}\mathbf{\tilde I}_{a,n}x_n(t)$, respectively. Then, the power consumption of EAP, denoted by $P_t$, can be given by
\begin{equation}\label{eq:P_t}
\begin{split}
P_t&\!=\!\Re \left\{\frac{1}{WT_S}\!\!\int\!\mathbb E\left\{\mathbf{i}_{a}(t)^H\mathbf{v}_{a}(t)\right\}dt\right\}
\\&\!=\!\frac{B}{2}\sum_{n=1}^{N}\mathbf{\tilde I}_{a,n}^H(\mathbf{\bar Q}\!+\!\mathbf{\bar Q}^H\!+\!\mathbf{\Delta Q}\!+\!\mathbf{\Delta Q}^H)\mathbf{\tilde I}_{a,n}.
\end{split}
\end{equation}
where $\Re\{\cdot\}$ denotes the operation of taking the real part.

To find a robust magnetic beamforming scheme which can minimize the power consumption of EAP while guaranteeing the sum rate and the SINR of each TU in the presence of estimation errors, the following problem, denoted by $\textbf{P1}$, is formulated:
 \begin{equation}\label{eq:P_1}
\begin{split}
&\textbf{P1:}\quad\mathop{\min }\limits_{\{\mathbf{\tilde I}_{a,n}\}} P_t
\\&{\text{ s}}{\text{.t}}{\text{. :}}~1).~C_S\ge C_{th};
\\&~~~~~~~2).~\gamma_n\ge \gamma_{th,n}, ~n=1,2,...,N;
\\&~~~~~~~3).~\|\mathbf{\Delta H}_n\|_F\leq\xi_n,~n=1,2,...,N;
\\&~~~~~~~4).~\|\mathbf{\Delta Q}\|_F\leq\xi_q.
\end{split}
\end{equation}
In \textbf{P1}, $C_S$ denotes the threshold of network sum rate and $\gamma_{th,n}$ represents the SINR threshold of the $n$-th TU. It can be observed that constraint 1) is non-convex. Also, due to the randomness and continuity of $\mathbf{\Delta H}_n$ and $\mathbf{\Delta Q}$, \textbf{P1} is with a non-convex objective function and infinite non-convex constraints. Thus, \textbf{P1} is difficult to be directly solved. In the following $\textbf{P1}$ is transformed into convex and solved.
\section{Robust Magnetic Beamforming Design}
We can observe from $\textbf{P1}$ that the objective function is highly non-convex due to the randomness and continuity of $\mathbf{\Delta Q}$.
To solve $\textbf{P1}$, we derive an upper-bound for $\mathbf{\tilde I}_{a,n}^H(\mathbf{\Delta Q}\!+\!\mathbf{\Delta Q}^H)\mathbf{\tilde I}_{a,n}$ as follows:
\begin{equation}\label{eq:UP}
\begin{split}
&\mathbf{\tilde I}_{a,n}^H(\mathbf{\Delta Q}\!+\!\mathbf{\Delta Q}^H)\mathbf{\tilde I}_{a,n}\!\mathop=^{\text{(a)}} \!\|\mathbf{\tilde I}_{a,n}^H(\mathbf{\Delta Q}\!+\!\mathbf{\Delta Q}^H)\mathbf{\tilde I}_{a,n}\|_F
\\&\mathop\leq^{\text{(b)}} \! \|\mathbf{\tilde I}_{a,n}^H\|_F\!\times\!\|\mathbf{\Delta Q}\!+\!\mathbf{\Delta Q}^H\|_F\!\times\!\|\mathbf{\tilde I}_{a,n}\|_F
\\&\mathop\leq^{\text{(c)}} \! \|\mathbf{\tilde I}_{a,n}\|_F^2\!\times\!\!\|\mathbf{\Delta Q}\|_F\!+\!\!\|\mathbf{\tilde I}_{a,n}\|_F^2\!\times\!\!\|\mathbf{\Delta Q}^H\|_F\!=\!2\xi_q\mathbf{\tilde I}_{a,n}^H\mathbf{\tilde I}_{a,n},
\end{split}
\end{equation}
where (a) is because $\mathbf{\tilde I}_{a,n}^H(\mathbf{\Delta Q}\!+\!\mathbf{\Delta Q}^H)\mathbf{\tilde I}_{a,n}$ is a real number, (b) results from the Cauchy-Schwarz inequality, and (c) results from the triangle inequality. Plugging \eqref{eq:UP} into \eqref{eq:P_t}, an upper-bound of $P_t$, denoted by $P_t^{U}$, can be given by
\begin{equation}\label{eq:P_t_UP}
\begin{split}
P_t\leq P_t^{U}=B\sum_{n=1}^{N}\mathbf{\tilde I}_{a,n}^H(\Re\{\mathbf{\bar Q}\}\!+\!\xi_q)\mathbf{\tilde I}_{a,n}.
\end{split}
\end{equation}
Based on the worst-case optimization criterion, $\textbf{P1}$ can be reformulated as the following optimization problem, denoted by $\textbf{P2}$:
 \begin{equation}\label{eq:P_2}
\begin{split}
&\textbf{P2:}\quad\mathop{\min }\limits_{\{\mathbf{\tilde I}_{a,n}\}} P_t^U
\\&{\text{ s}}{\text{.t}}{\text{. :}}~1), 2), \text{and}~3)~\text{in}~\textbf{P1}.
\end{split}
\end{equation}
\begin{figure*}
\begin{equation}\label{eq:Equivalent_C2}
\mathbf{\Gamma}_n(\{\mathbf{X}_n\}, a_n,\kappa_n)\!=\!\left[\begin{array}{cc}
\mathbf{X}_n-\kappa_n\sum_{u\ne n}^N\mathbf{X}_u\!+\!a_n \mathbf{I} & (\mathbf{X}_n-\kappa_n\sum_{u\ne n}^N\mathbf{X}_u) \mathbf{\bar H}_n^H\\
 \mathbf{\bar H}_n(\mathbf{X}_n-\kappa_n\sum_{u\ne n}^N\mathbf{X}_u)^H & \mathbf{\bar H}_n(\mathbf{X}_n-\kappa_n\sum_{u\ne n}^N\mathbf{X}_u)\mathbf{\bar H}_n^H-a_n\xi_n-\kappa_nN_0
\end{array}\right] \succeq 0.
\end{equation}
\hrulefill
\vspace{-10pt}
\end{figure*}To cope with the non-convex sum rate constraint $C_S\!\ge\! C_{th}$, we introduce auxiliary coefficient $\kappa_n\!=\!\max\{\gamma_{th,n},2^{[C_{th}/(BN)]}\!-\!1]\}$. If $\gamma_n\ge\kappa_n$, both 1) and 2) of \textbf{P2} can be satisfied. With $\kappa_n$, \textbf{P2} is further transformed into the following problem, denoted by $\textbf{P3}$:
 \begin{equation}\label{eq:P_3}
\begin{split}
&\textbf{P3:}\quad\mathop{\min }\limits_{\{\mathbf{\tilde I}_{a,n}\}} P_t^U
\\&{\text{ s}}{\text{.t}}{\text{. :}}~1).~\gamma_n\ge \kappa_{n}, ~n=1,2,...,N;
\\&~~~~~~~2).~\|\mathbf{\Delta H}_n\|_F\leq\xi_n,~n=1,2,...,N.
\end{split}
\end{equation}
However, \textbf{P3} is still with infinite constraints due to the randomness and continuity of $\mathbf{\Delta H}_n$. Next we exploit the S-procedure~\cite{Convex_Optimization} to reformulate it. According to the S-procedure, for $\mathbf{x}\in\mathbb C^{K}$, $\mathbf{A}_1\in\mathbb C^{K\times K}$, $\mathbf{A}_2\in\mathbb C^{K\times K}$, $\mathbf{b}_1\in\mathbb C^{K}$, $\mathbf{b}_2\in\mathbb C^{K}$, $c_1\in\mathbb R$, and $c_2\in\mathbb R$, if $\mathbf{x}^H\mathbf{A}_1\mathbf{x}^H+2\mathcal R\{\mathbf{b}_1\mathbf{x}\}+c1\leq0$, then $\mathbf{x}^H\mathbf{A}_2\mathbf{x}^H+2\mathcal R\{\mathbf{b}_2\mathbf{x}\}+c2\leq0$ holds if and only if there exists $a \ge 0$ such that
\begin{equation}\label{eq:SP}
\left[\begin{array}{cccc}
\mathbf{A}_2& \mathbf{b}_2 \\
\mathbf{b}_2^H & c_2
\end{array}\right]\preceq
a\left[\begin{array}{cccc}
\mathbf{A}_1& \mathbf{b}_1 \\
\mathbf{b}_1^H & c_1
\end{array}\right].
\end{equation}
We denote $\mathbf{X}_n=\mathbf{\tilde I}_{a,n}^H\mathbf{\tilde I}_{a,n}$. Based on the S-procedure and the constraint 2) of \textbf{P3}, the constraint 1) of \textbf{P3} holds if and only if there exists $a_n$, such that \eqref{eq:Equivalent_C2} holds.
Then, \textbf{P3} can be further reformulated as
 \begin{equation}\label{eq:P_4}
\begin{split}
&\textbf{P4:}\quad\mathop{\min }\limits_{\{\mathbf{X}_n\}} \sum_{n=1}^{N}{\rm Tr}\left\{\left(\Re\{\mathbf{\bar Q}\}\!+\!\xi_q\right) \mathbf{X}_n\right\}
\\&{\text{ s}}{\text{.t}}{\text{. :}}~1).~\mathbf{\Gamma}_n(\{\mathbf{X}_n\}, a_n,\kappa_n)\succeq 0, ~n=1,2,...,N;
\\&~~~~~~~2).~{\rm Rank}(\mathbf{X}_n)=1,~n=1,2,...,N;
\\&~~~~~~~3).~\mathbf{X}_n\succeq 0,~n=1,2,...,N,
\end{split}
\end{equation}
where ${\rm Rank}(\cdot)$ and ${\rm Tr}\{\cdot\}$ denote the operations of taking the rank and trace, respectively. Using semidefinite relaxation (SDR) and dropping the non-convex rank-one constraint, the relaxed problem of \textbf{P4} is a semidefinite programming (SDP), which can be solved with convex tools such as CVX. Also, many existing methods, such as Gaussian randomization~\cite{Gang_Yang_TSP,Yixin_Zhang_TSP,Robust_Beamforming_2} can be used to obtain a rank-one solution to \textbf{P4}. Due to page limitation, we omit the lengthy and repetitive procedures of finding a rank-one solution to \textbf{P4}. We denote by $\mathbf{X}_n^*$ the rank-one solution to \textbf{P4}. Then, the proposed robust beamforming vector, denoted by $\mathbf{\tilde I}_a^*$, can be obtained as $\mathbf{\tilde I}_a^*=\sqrt{\epsilon_n}\mathbf{e}_n$, where $\epsilon_n$ and $\mathbf{e}_n$ denote the eigenvalue and eigenvector of $\mathbf{X}_n^*$, respectively.
\section{Numerical Simulations}
In this section, numerical simulations are conducted to show the performances of the proposed robust magnetic beamforming scheme.
The transmission medium is set to soil, whose relative permeability is 1, the relative permittivity is 7, and the conductivity is $10^{-1}$~S/m.
The transmission bandwidth is set to $B=0.1$~kHz and the operating frequency is set to $f_c=1$~MHz. The coil radii of AP and BDs are set to $80$~cm and $30$~cm, respectively. The number of turns of EAP coils and TU coils are set to $20$ and $12$, respectively. We set the number of AP coils to $K=5$ and the number of TUs to $N=2$. The TUs' locations are randomly generated within a radius of 15~m below the EAP. The estimation error for $\mathbf{Q}$ and $\mathbf{H}_n$ are characterized by the norm-bounded model with $\xi_q=g\|\mathbf{\bar Q}\|_F$ and $\xi_n=g\|\mathbf{\bar H}_n\|_F$, where $g$ is the introduced error parameter. For each value of $g$, we do $2\times10^3$ tests to calculate the outage probability.
The outage probability of the $n$-th TU is defined as $P_{\text{out},n}\!=\!\text{Pr}\{\gamma_n\!<\!\gamma_{th,n}\}$. If a user outage occurs, we determine that a network outage happens. Therefore, the network outage probability is defined as $P_{\text{out}, Net} = 1\!-\!(\prod_{n=1}^{N} P_{\text{out}, n})$.
The effective throughput is defined as the average sum rate conditioned on the network being in non-outage states.
We compare the proposed robust beamforming scheme with two non-robust schemes. The first one is referred to as non-robust beamforming, which assumes that perfect CSI is known and obtained by solving $\textbf{P1}$ with $\mathbf{\Delta Q}=\mathbf{0}$ and $\mathbf{\Delta H}_n=\mathbf{0}$. The second one is conventional minimum mean squared error (MMSE) scheme, which also assumes that perfect CSI is known and obtained by solving the following problem with $\mathbf{\Delta Q}=\mathbf{0}$ and $\mathbf{\Delta H}_n=\mathbf{0}$:
 \begin{equation}\label{eq:P_5}
\begin{split}
&\textbf{P5-MMSE:}\quad\mathop{\min }\limits_{\{\alpha_n\}} P_t
\\&{\text{ s}}{\text{.t}}{\text{. :}}~1).~C_S\ge C_{th};
\\&~~~~~~~2).~\gamma_n\ge \gamma_{th,n}, ~n=1,2,...,N;
\\&~~~~~~~3).~\mathbf{\tilde I}_{a,n}=\alpha_n\mathbf{W}_n, ~n=1,2,...,N.
\end{split}
\end{equation}
where $\mathbf{W}_n\!\!=\!\!\mathbf{H}^H(\mathbf{H}\mathbf{H}^H\!\!+\!\!N_0\mathbf{I})\mathbf{o}_n^H$ with $\mathbf{H}\!\!=\!\![\mathbf{H}_1^H,...,\mathbf{H}_N^H]^H$.
 \begin{figure}
\centering
\includegraphics[width=0.4\textwidth]{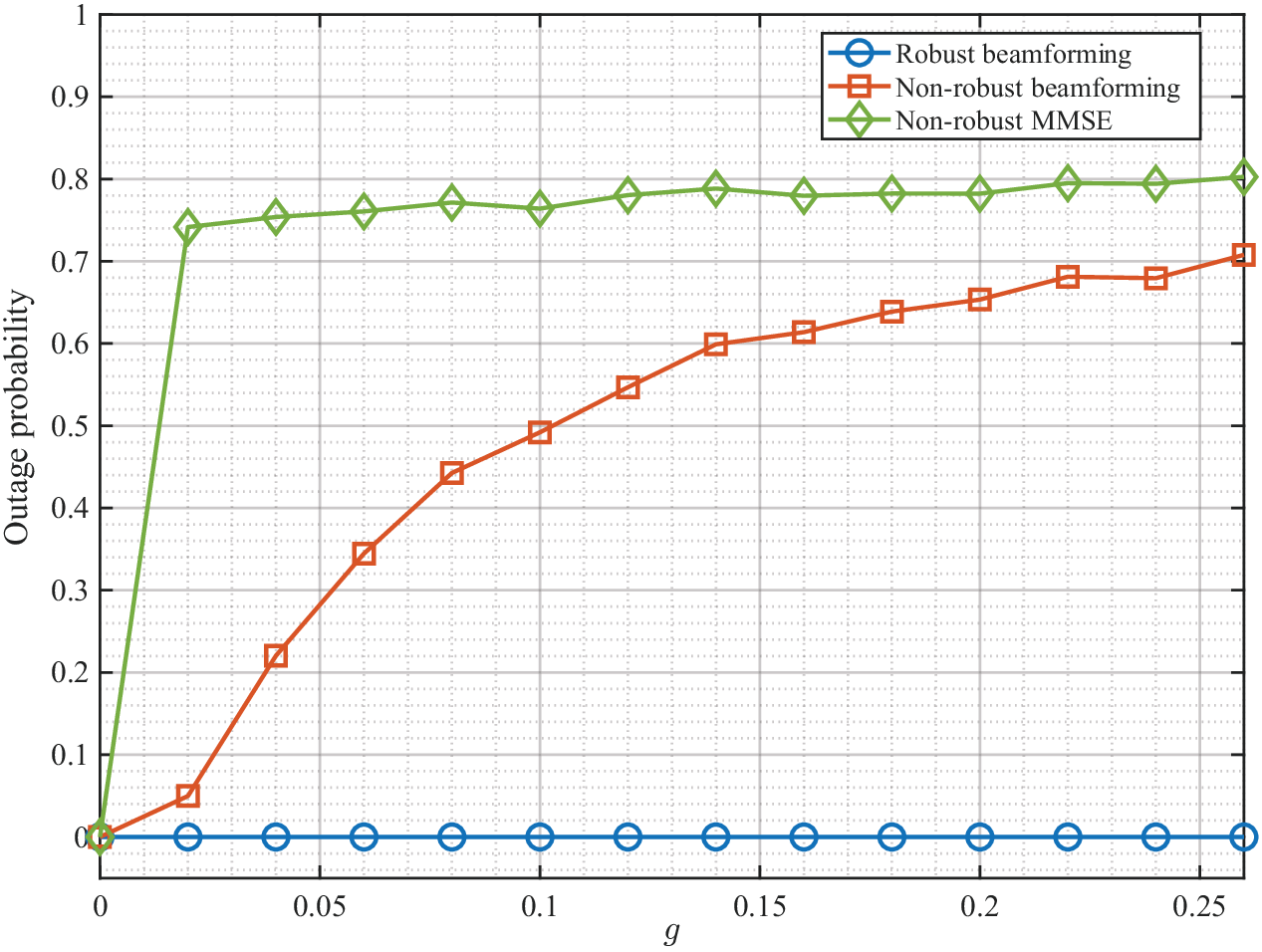}
\caption{Network outage probabilities of the proposed robust beamforming, non-robust beamforming, and the non-robust MMSE.}\label{fig:Outage_Probability}
\vspace{-15pt}
\end{figure}

Figure~\ref{fig:Outage_Probability} shows the network outage probabilities of the proposed robust beamforming, non-robust beamforming, and the non-robust MMSE. As shown in Fig.~\ref{fig:Outage_Probability}, when the accuracy of estimation is low, the outage probabilities of the non-robust beamforming and non-robust MMSE are very high. For example, when $g=0.25$, the outage probabilities of the non-robust beamforming and non-robust MMSE are 0.7 and 0.8, respectively. In addition, it is shown that non-robust MMSE is very sensitive to estimation error. When $g=0.02$, the network outage probability of non-robust MMSE is 0.74.
Compared with the two schemes, the proposed robust beamforming achieves non-outage, which shows that the proposed robust scheme has good communication reliability and robustness to the measurement and estimation errors.

\begin{figure}
\centering
\includegraphics[width=0.395\textwidth]{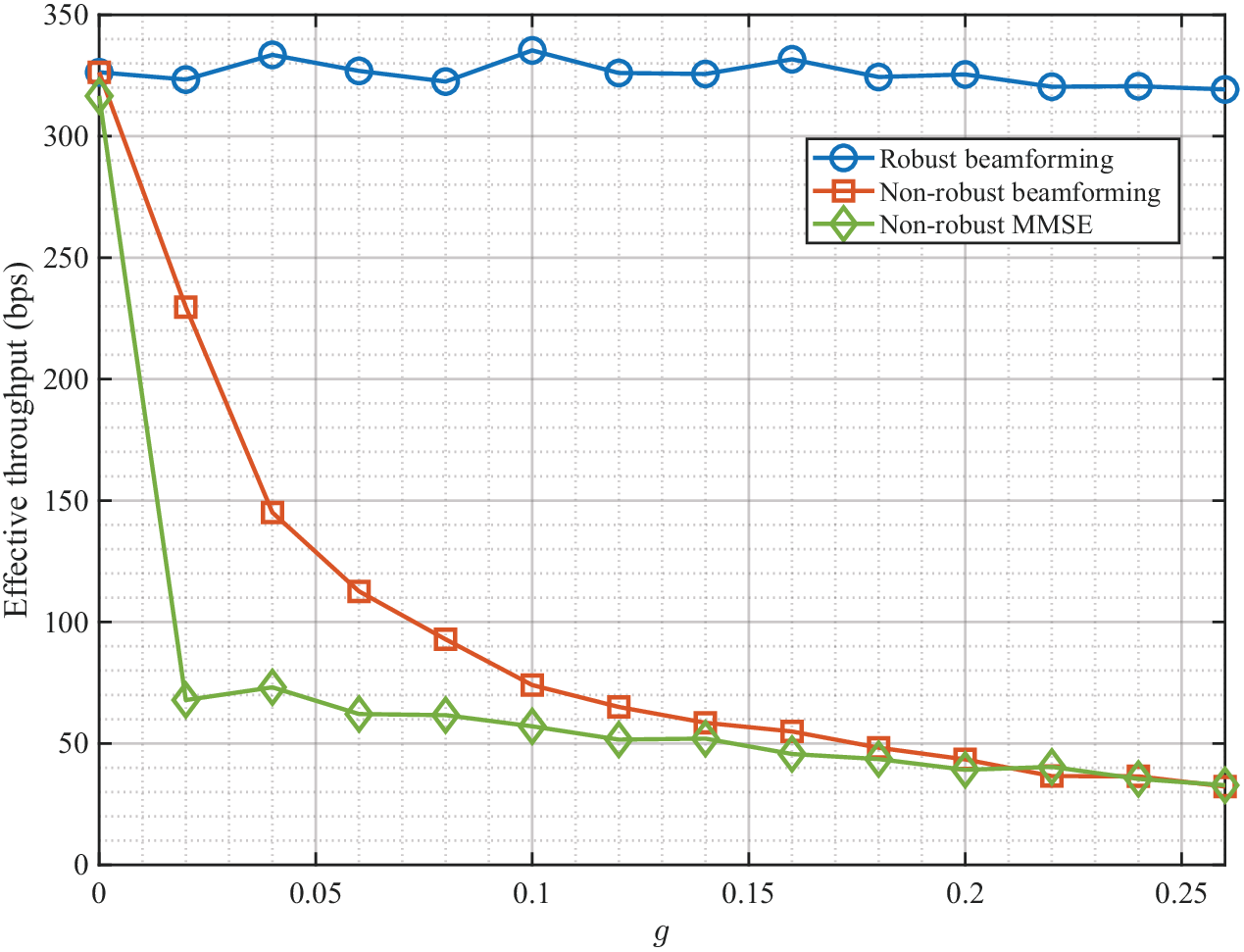}
\caption{The effective throughputs of the proposed robust beamforming, non-robust beamforming, and the non-robust MMSE.}\label{fig:Effective_Throughput}
\vspace{-15pt}
\end{figure}

The effective throughputs of the three schemes are shown in Fig.~\ref{fig:Effective_Throughput}. We can observe that the effective throughputs of non-robust schemes, especially non-robust MMSE, decrease quickly as $g$ increases. When the accuracy of estimation is low, the effective throughputs of the non-robust beamforming and non-robust MMSE are similar. Compared with the two schemes, the proposed robust beamforming has stable and high effective throughput, which verifies the reliability and robustness of the proposed robust beamforming.
\section{Conclusion}
To enhance the reliability and effective throughput of MI communication in post-disaster rescues, in this paper we investigated the problem of robust magnetic beamforming for MI based multi-user underground emergency communication under imperfect CSI, which aims at minimizing transmit power while simultaneously guaranteeing the network sum rate and the SINR requirements for TUs. By leveraging the worst-case optimization criterion and the S-procedure, the original non-convex problem was transformed into a tractable convex form and solved. Numerical results showed that as compared to non-robust approaches, our method exhibits superior performance in terms of reliability and effective throughput, making it a promising solution for practical underground emergency communication scenarios.
\bibliographystyle{IEEEbib}
\bibliography{ref}
\end{document}